\documentstyle[11pt,newpasp,twoside,epsf]{article}
\def\edcomment#1{\iffalse\marginpar{\raggedright\sl#1\/}\else\relax\fi}
\reversemarginpar

\begin{document}
\title{The dark matter halos of spheroidal galaxies and clusters of galaxies}
 \author{Tommaso Treu}
\affil{California Institute of Technology, Astronomy 105-24, Pasadena CA 91125; present address: University of California at Los Angeles, Astronomy \& Astrophysics, Los Angeles, CA 90095}
\author{L\'eon V.~E. Koopmans}
\affil{Space Telescope Science Institute, 3700 San Martin Dr, Baltimore MD 21218}
 \author{David J. Sand, Graham P. Smith, Richard S. Ellis}
\affil{California Institute of Technology, Astronomy 105-24, Pasadena CA 91125}

\begin{abstract}
We describe the first results from two observational projects aimed at
measuring the amount and spatial distribution of dark matter in
distant early-type galaxies (E/S0s) and clusters of galaxies. At
the galaxy scale, the Lenses Structure and Dynamics (LSD) Survey is
gathering kinematic data for distant (up to $z\sim1$) E/S0s that are
gravitational lenses. A joint lensing and dynamical analysis
constrains the fraction of dark matter within the Einstein radius, the
mass-to-light ratio of the stellar component, and the total slope of
the mass density profile. These properties and their evolution with
redshift are briefly discussed in terms of the formation and evolution
of E/S0
galaxies and measurement of the Hubble Constant from gravitational
time delay systems. At the cluster scale -- after careful removal of the
stellar component with a joint lensing and dynamical analysis --
systems with giant radial arcs can be used to measure precisely the
inner slope of the dark matter halo. An HST search for radial arcs and
the analysis of a first sample are briefly discussed in terms of the
universal dark matter halos predicted by CDM simulations.

\end{abstract}

\section{Introduction}

Decades after the discovery of dark matter around spiral galaxies
little is known about dark matter in early-type galaxies. Mass tracers
at large radii (such as stellar kinematics, kinematics of globular
clusters and of planetary nebulae, and X-ray halos) generally indicate
that a constant mass-to-light ratio cannot reproduce the observations,
although there are typically large uncertainties and a wide variety of
behavior is seen. The main source of uncertainty in interpreting
kinematic measures is that the derived mass profile depends on the
assumed orbital structure. This is commonly referred to as the
mass-anisotropy degeneracy.

The situation is even more uncertain outside the local Universe,
because traditional kinematic tracers at large radii are generally not
feasible at cosmological distances. However, additional constraints
can be gathered by looking at early-type galaxies that are
gravitational lenses. The configuration of multiple images provides
information on the mass distribution of the lens; first and foremost a
robust measurement of the mass enclosed within the Einstein Radius
(typically larger than the effective radius), {\it independent} of its
kinematic status.

The Lenses Structure and Dynamics (LSD) Survey (Koopmans \& Treu 2002,
2003; Treu \& Koopmans 2002a,2003; hereafter KT) is obtaining internal
kinematics for a sample of of 11 relatively isolated E/S0 lens
galaxies at $z=0.04-1.01$. A joint lensing and dynamical analysis is
used to break the mass-anisotropy degeneracy and determine the amount
and distribution of luminous and dark matter. The first results from
this project are briefly summarized and discussed in Section~2.

At the cluster scale, the presence of dark matter has been known for a
long time (Zwicky 1937). The distribution of dark matter has been
probed over a wide range of scales using X-ray analysis, dynamical
studies and gravitational lensing, showing that dark matter dominates
in mass. The wealth of mass tracers in clusters -- in particular the
existence of giant radial arcs -- makes them an ideal laboratory to
test the existence of universal mass density profiles predicted by
cold dark matter (CDM) simulations. Combining a lensing analysis of
clusters with giant arcs with a dynamical analysis of the kinematics
of the brightest cluster galaxy (BCG), we can measure very accurately
the mass density profile within $\sim100$ kpc, disentangle the stellar
and dark matter, and pinpoint the inner logarithmic slope of the dark
matter halo. This has motivated a search of suitable systems with a
central dominant BCG and giant arcs through the entire HST-WFPC2
archive (Sand et al.\ 2004, in preparation), and the spectroscopic
follow-up at Keck (Sand, Treu \& Ellis 2002; Sand et al.\ 2003). The
first results from this project are briefly summarized and discussed
in Section 3.

\section{Early-type galaxies: dark halos and cosmic evolution}
\label{sec:eso}

\begin{figure}
\plottwo{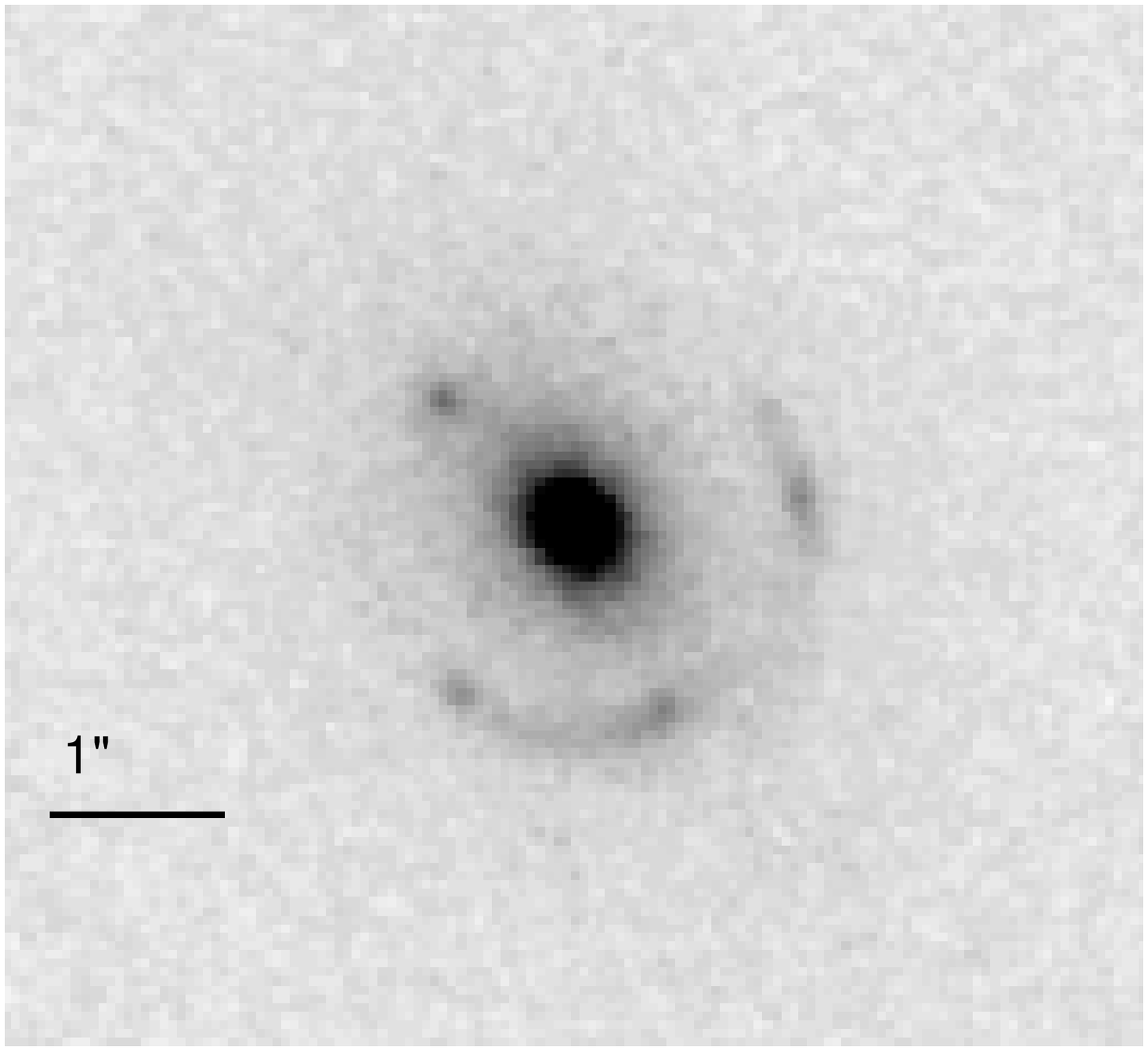}{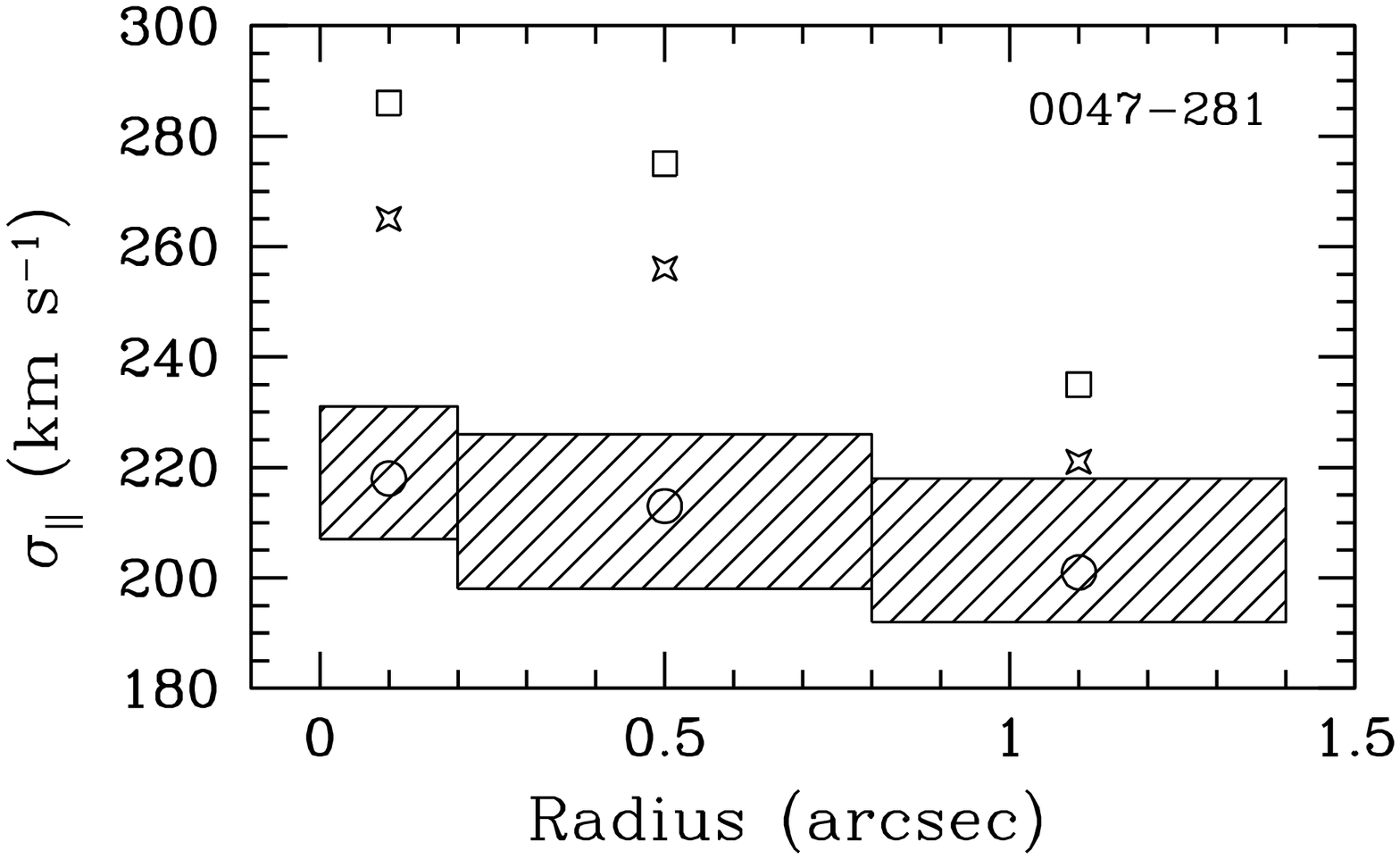}
\caption{Left: HST image of 0047$-$281 at $z=0.485$. Right: velocity
dispersion profile of 0047$-$281 along the major axis. The box height
indicates the 68\% measurement error, whereas the box width indicates
the spectroscopic aperture. The open squares are the corresponding
values for an isotropic constant $M/L$ model, which is rejected by the
data. See Koopmans \& Treu (2003) for details.}
\end{figure}

Figure~1 illustrates via the lens 0047-281 ($z=0.485$; Warren et
al. 1996) the 
data we are collecting as part of the LSD Survey. Archival HST imaging
data (mostly from the CASTLeS collaboration) are combined with 
stellar kinematic data obtained using ESI at the Keck-II
Telescope. In excellent conditions (seeing $\sim0.6''$) it is possible
to obtain a spatially resolved stellar velocity dispersion profile
(hatched boxes in the right panel) out to the Einstein Radius
(typically 1-2 effective radii, 1-2 arcseconds). A joint lensing and
dynamical analysis is used to constrain the free parameters of a
family of two-component mass models. One component represents the
stellar mass and follows the HST luminosity profile scaled by a
stellar mass-to-light ratio, the other component represents the dark
matter halo and is modeled as a generalized Navarro, Frenk \& White
(1997) profile (see TK for details).

The main results from analysis of the first two objects (0047-281 and
MG2016) are:
 
\smallskip \noindent 1. Extended dark matter halos are detected in
high redshift E/S0 galaxies at high significance
($>10$~$\sigma$); dark matter contributes 50--75 \% of the total mass
within the Einstein radius. 

\smallskip\noindent 2. The evolution of the stellar mass-to-light
ratio with redshift, obtained from a two-component dynamical model,
agrees with the independent estimate obtained by studying the
evolution of the Fundamental Plane with redshift (Treu et al.\ 2002;
van Dokkum \& Ellis, Gebhardt et al.\ 2003). This is consistent with
no structural or dynamical evolution of E/S0s between $z=1$ and today.

\smallskip\noindent 3. The effective slope (i.e. the average slope
measured between roughly the effective and Einstein radius) of the
{\sl total} mass distribution is very close to isothermal,
i.~e. $\rho\propto r^{-\gamma'}$ with $\gamma'\sim2.0$ (to within
5\%), which we interpret as a possible indication of (incomplete)
violent relaxation. This result, if confirmed by a larger sample,
indicates that already at $z\sim1$ there is a mechanism by which dark
and luminous component conspire with each other to produce almost
perfectly flat rotation curves, while still preserving their spatial
segregation.

Clearly, analysis of the rest of the sample is essential to confirm
and extend these findings.  For example, an open question -- which we
intend to address and has profound cosmological implications -- is
understanding what is the distribution of the effective slope of total
mass density profile. An intrinsic scatter is expected on the basis of
local studies (e.g. Gerhard et al.\ 2001). It is necessary to quantify
this scatter in order to estimate the uncertainties in the
determination of the cosmological parameters from lens statistics, of
velocity dispersion of early-type galaxies from image separation, and
of the Hubble Constant from gravitational time delays.  The Hubble
Constant is highly sensitive to the slope of the mass density profile,
and some external information is needed if one wants to measure it
with precision better than a few tens of per cent. In an effort
parallel to the LSD Survey, an enlarged collaboration is targeting
lenses with known time-delays for spectroscopic follow-up with Keck
and VLT. Although this is generally complicated by the presence of
bright QSO images outshining the lens galaxy, analysis of the first
two systems (Treu \& Koopmans 2002a; Koopmans et al.\ 2003) confirms
that with suitable kinematic data it is possible to measure H$_0$ at
$\sim$15\% accuracy from each individual system, {\it including the
uncertainty on the mass density profile}. We are collecting data for
more systems, with the goal of measuring H$_0$ from time-delays with
precision comparable to the HST Key Project (Freedman et al.\ 2001).

\section{Galaxy clusters: the inner density profile and the CDM
  ``cusp'' problem} 

As part of our study of the inner regions of clusters of galaxies we
have measured arc redshifts and BCG stellar velocity dispersions for 6
clusters of galaxies (Figure~2; Sand et al.\ 2002, 2003). Three
systems (upper panels) are selected to have both a radial and a
tangential gravitational arc, while three systems (lower panels) are
selected to have only tangential arcs. The location of the symetry
break in a radial arc constrains the derivative of the enclosed mass
and therefore the radial arc systems provide very tight constraints on
the shape of the mass density profile. However, the radial critical
line disappears when the total mass density profile becomes steeper
then $r^{-2}$. Therefore a radial arc selected sample may possibly be
biased toward flatter mass profiles. For this reason we introduced the
tangential arc only systems as a control sample.

\begin{figure}[h!] 
\plotone{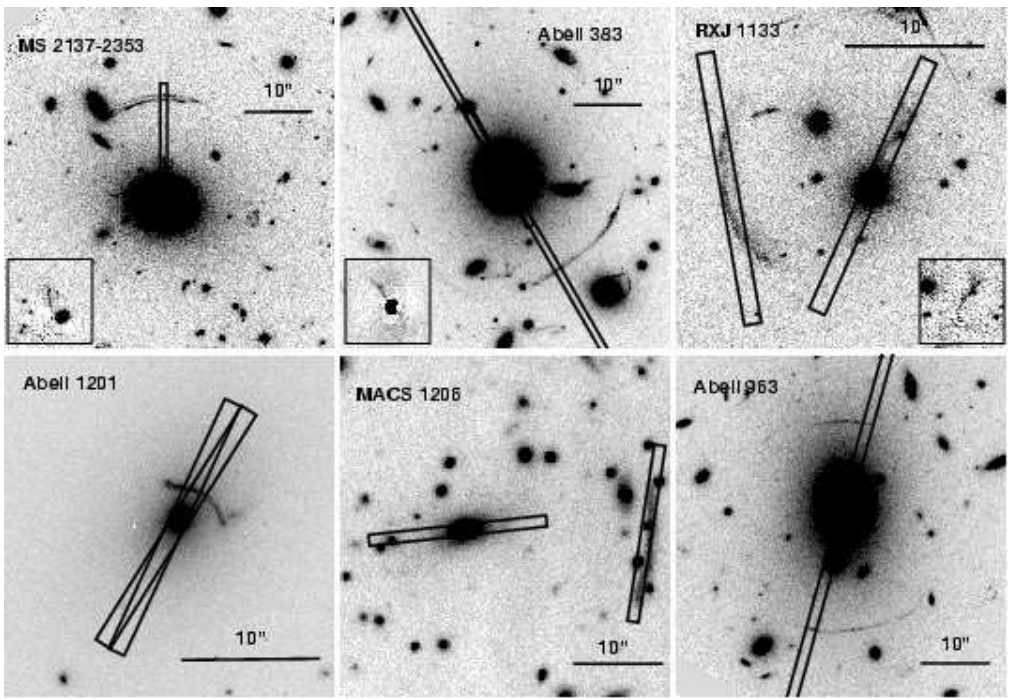}
\caption{Images of the six clusters studied by Sand et al.\ (2003)}
\end{figure}

A joint lensing and dynamical analysis is performed, fitting a family
of two component mass models to the data (see Sand et al. 2003 for
details). Specifically, the dark matter mass component is parametrized
as a generalized NFW profile (Navarro, Frenk \& White 1996,1997; Moore et
al.\ 1998), i.e. the dark matter density goes as $r^{-\beta}$ at small
radii and $r^{-3}$ at large radii. Numerical
simulations predict $\beta$ in the range 1-1.5 (hereafter,
respectively, NFW and Moore slopes), although numerical issues are
still being debated (see Sand et al. 2003 and references therein).

\begin{figure}[h!] 
\plottwo{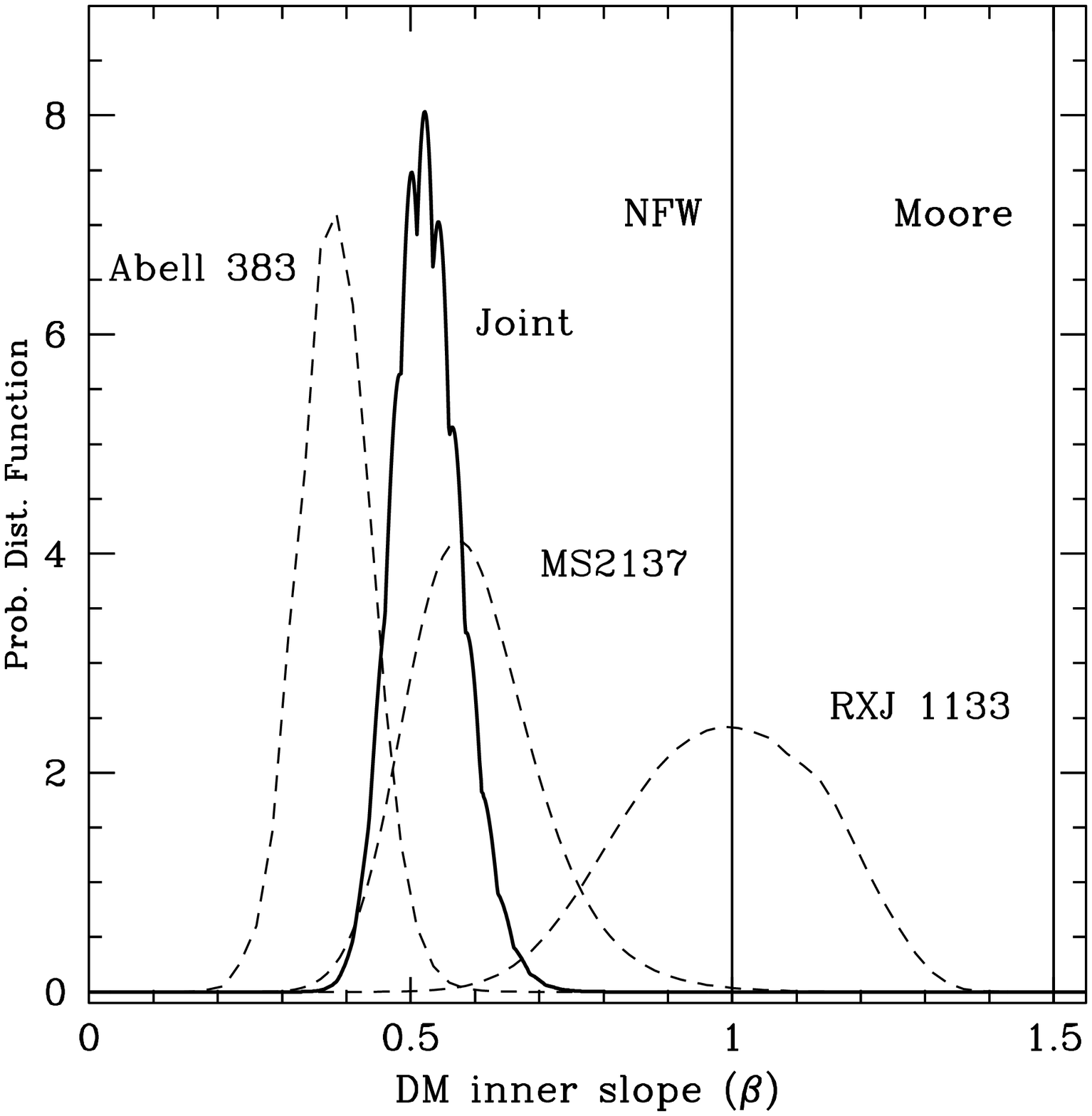}{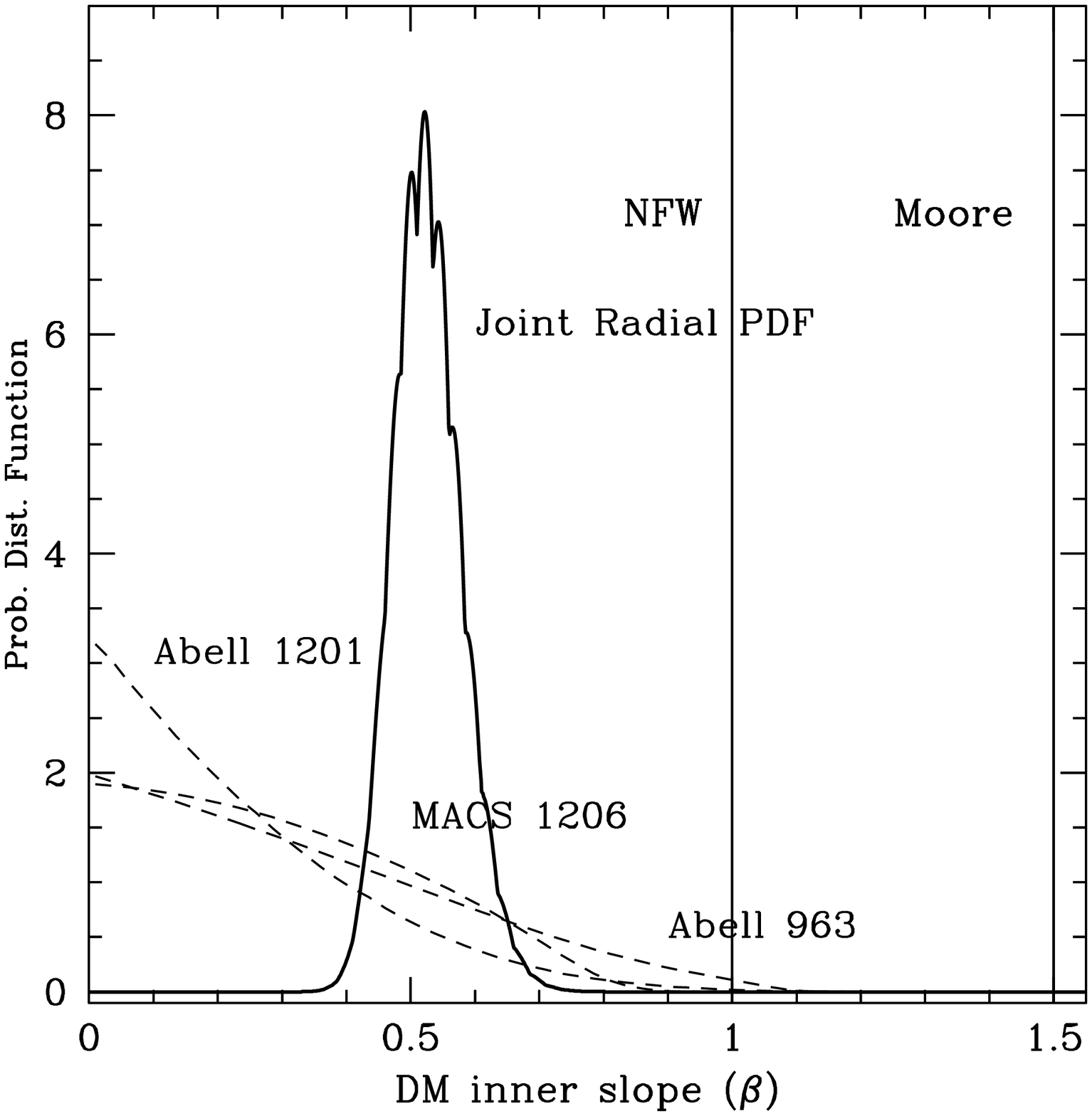}
\caption{Left: Probability distribution function of the DM inner
density slope, $\beta$, for the three radial arc clusters. Right:
Probability distribution function of the DM inner density slope,
$\beta$, for the tangential arc sample. Also plotted is the joint PDF
for the radial arc sample and the tangential arc sample.  (from Sand
et al.\ 2003) }
\end{figure}

Figure~3 shows the posterior probability density for $\beta$ for the
radial (left) and tangential (right) arc samples. The radial arc
sample indicates generally flat inner slopes, the average slope being
inconsistent with the CDM ``cusps'' at more than 99.99\% CL. However,
intrinsic scatter is detected and individual systems can be as steep
as NFW (RXJ1133). The tangential arc systems also confirm flatter
slopes than $\beta=1$, indicating that the sample with radial arcs is
unbiased.

The implications of this discrepancy could be profound. Either the
CDM-only simulations are insufficiently accurate and/or contain
insufficient physics to describe faithfully the inner
regions of clusters. For example, baryons could play an important
dynamical role driving energy and angular momentum out of the cluster
core through mechanisms such as dynamical friction (e.g. El-Zant et
al.\ 2003). Alternatively, the DM could be partially self-interacting,
or manifest some interaction with the baryons (Spergel \& Steinhardt
2000). Either outcome would be very exciting. On the one hand,
numerical simulations are being developed which will permit the
realistic inclusion of baryons.  Finding mechanisms that satisfy our
observational constraints will be invaluable in understanding the
complex physics of star formation and dissipative processes. And of
course, there is the possibility that there is some fundamental flaw
in CDM which will necessitate considering more exotic scenarios, like
self-interacting DM.

We are currently looking to expand our sample, drawing from the
candidate radial arc systems identified from the HST-WFPC2 archive
(Sand et al.\ 2004). This would improve our determination of the mean
of $\beta$ and also provide a first measure of the intrinsic scatter,
which is essential for any meaningful comparison with simulations,
especially those that include baryons.

\section{Acknowledgments}

TT thanks the organizers for this pleasant and exciting meeting.  TT
thanks Eric Agol, Stefano Borgani, Amr El-Zant, Raul Jimenez, Masataka
Fukugita, Jean-Paul Kneib, Ben Moore, Joel Primack, Paul Schechter,
Jerry Sellwood and Simon White for stimulating conversations, during
and around the Sydney Meeting. TT gratefully acknowledges support from
an AAS travel grant and from HST proposals AR-09222, AR-09527,
AR-09960, provided by NASA through a grant from STScI, which is
operated by AURA, Incorporated, under NASA contract NAS5-26555.

\end{document}